\documentclass{PHYEAUTH}
\usepackage{natbib}
\usepackage{graphicx}
\usepackage{amsmath}
\usepackage{amssymb}

\begin{document}

\begin{frontmatter}

\title{Scattering mechanisms and Boltzmann transport in graphene}

\author[address1]{Shaffique Adam\thanksref{thank1}},
\author[address1]{E. H. Hwang}
and
\author[address1]{S. Das Sarma}

\address[address1]{Condensed Matter Theory Center, Department of Physics,
University of Maryland, College Park, MD 20742-4111, USA}

\thanks[thank1]{
Corresponding author.  E-mail: adam1@umd.edu}

\date{7 July 2007}
\begin{abstract}
Different scattering mechanisms in graphene are explored and
conductivity is calculated within the Boltzmann transport theory.  We
provide results for short-range scattering using the Random
Phase Approximation for electron screening, as well as analytical
expressions for the dependence of conductivity on the dielectric
constant of the substrate.  We further examine the effect of ripples
on the transport using a surface roughness model developed for
semiconductor heterostructures.  We find that close to the Dirac point,
$\sigma \sim n^\beta$, where $\beta=1,0,-2$ for Coulomb, short-range
and surface roughness respectively; implying that Coulomb scattering 
dominates over both short-range and surface roughness scattering 
at low density.    
\end{abstract}

\end{frontmatter}

\section{Introduction}

Over the last two years, there has been a proliferation in theoretical
and experimental interest in graphene.  Graphene is a two dimensional
sheet of Carbon atoms arranged in a honey-comb lattice, or
alternatively, a mono-atomic sheet of carbon that has been cleaved from
bulk graphite onto a semiconductor substrate.  Graphene has generated
much excitement as theorists and experimentalists explore both the
familiar and peculiar features of this newly discovered condensed matter
system.  Perhaps the most curious property of graphene was the ``minimum
conductivity'' puzzle, where a na\"ive examination of experimental results
would suggest that graphene had a finite conductivity even as the
density of carriers vanishes.  This ``carrier-free'' universal minimum
conductivity has been the subject of much speculation in the recent
literature.  However, there is now mounting
evidence~\cite{kn:tan2007,kn:cho2007} suggesting a more sensible
alternative -- charged impurities that are invariably present in the
substrate create an inhomogeneous potential
landscape~\cite{kn:hwang2006c}.  At large external gate voltage $V_g =
n/\alpha \gg n_{imp}/\alpha$, where $n_{imp}$ is the density of charged
substrate impurities and $\alpha$ is related to the capacitance of the
substrate, the potential fluctuations can be ignored, but at low
density, they locally dope the graphene sheet breaking the system into
puddles of electrons and holes.  This spatial inhomogeneity has been
directly observed in recent experiments~\cite{kn:martin2007}.  We
observe that the charged impurities therefore have a dual role: first,
the induced graphene carrier density needs to be determined by the
screened charged impurity potential, and second, the conductivity is 
determined by charged impurity scattering.
         
Unlike gapped $2$D systems -- where much of the same physics is at play
--, in graphene the electrons and holes are both conducting.  So whereas
in semiconductor systems one might have a percolation transition between
a conducting and insulating state~\cite{kn:dassarma2005b}, graphene
transport, through this percolating network of electron and hole
puddles, is a transition between two conducting states.  One also finds
that the system self-averages at length scales of the correlation length
$L_\xi$, where $L_\xi \ll L$ and $L \sim 1 ~\mu \rm{m}$ is the typical
system size.  This then allows us to calculate the transport properties
using the {\it rms} carrier density $n^*$ induced by the potential
fluctuations (for details, see Ref.\cite{kn:adam2007a}).  This
mean-field theory is valid so long as $n^*$ is sufficiently large so
that the percolation transport is not critical -- a condition we believe
holds true for experiments currently being done on bulk graphene
samples. Ref.\cite{kn:cheianov2007} recently calculated 
the modification of the critical exponents due to the tunneling
between the electron and hole puddles. 

% The modification of the critical exponents due to the small
% tunneling between the electron and hole puddles was performed recently
% in Ref.\cite{kn:cheianov2007}, an effect that we ignore in our analysis.

In the present work, we examine two other types of scattering that may
be present in graphene, namely, short-range scattering (arising from
defects or dislocations in the Carbon lattice) and scattering from
ripples.  Both of these possible additional sources of scattering have
been observed in recent surface probe measurements of
graphene~\cite{kn:ishigami2007}. These experiments, however, also showed
that graphene sheets are relatively defect free and the ripples,
although present, are very small, where the average ripple height 
$\Delta = 1.9~\AA$, while the in-plane correlation length $\xi =
320~\AA$.  These numbers cast serious doubt over whether the electron
scattering off graphene ripples could have any observable effect on the
transport properties of graphene.  We observe that diffraction
experiments performed on suspended graphene sheets show ripples of about
the same height but with an order of magnitude smaller in-plane
correlation length~\cite{kn:meyer2007} further suggesting that the
ripples of graphene on a substrate is largely determined by the roughness
of that substrate.

For our purposes we reserve the term ``long-range'' scattering for
Coulomb impurities and use the two phrases interchangeably.  We consider
all delta-correlated disorder (i.e. Gaussian white noise potentials) to
be ``short-range'' scattering.  Other definitions have been employed in
the literature where white-noise disorder is sometimes described as
long-range if it preserves the graphene valley degeneracy and
short-range only if the disorder breaks this symmetry.  In all of
our discussion below we shall assume that both spin and valley
degeneracy is always preserved. 

Short-range scattering has been considered previously by many groups
including Refs.\cite{kn:ando2006,kn:aleiner2006,kn:nomura2007} using
various approximation schemes.  Here, we present results for the
screened short-range scattering using the Random Phase Approximation
(RPA) for electron screening~\cite{kn:hwang2006b} 
which allows us to extract the dependence of the conductivity
on the interaction parameter $r_s$, which can be tuned by 
using substrates of different dielectric constants.

% Using our formalism,
% we can extract the dependence of the conductivity $\sigma$ on the
% interaction parameter $r_s$, which in graphene experiments can be tuned
% by using substrates of different dielectric constants.

Coulomb scattering was considered by Ref.~\cite{kn:nomura2007} in the
completely screened approximation valid only in the limit $r_s
\rightarrow \infty$ and numerically in 
Ref.~\cite{kn:ando2006,kn:cheianov2006,kn:hwang2006c}.  To compare 
the different scattering mechanisms with
each other and to provide the framework for our calculation, we
reproduce here the analytic Coulomb scattering conductivity result that
was previously reported in Ref.~\cite{kn:adam2007a}.

The scattering of electrons off ripples has been of recent theoretical
interest.  Despite the magnitude of the ripples being quite small, it
is still believed to be responsible for the suppression of weak
localization~\cite{kn:morozov2006}, as well as causing changes to the
local chemical potential~\cite{kn:neto2007} and modification of the
transport properties~\cite{kn:katsnelson2007}.  To provide an estimate
for the contributions of static ripples to graphene transport
properties, we use a surface roughness model that was developed to
model interface potential inhomogeneities in semiconductor
heterojunction systems.  We do not attempt here to provide a rigorous
microscopic justification for applying this surface roughness
model~\cite{kn:ando1982} to graphene, although given its success in
modeling surface inhomogeneities in Si-MOSFETs, it provides a good
starting point to model the matrix element for electrons scattering
off the potential inhomogeneity caused by ripples.  Whereas the formal
validity of this model has yet to be established, we are persuaded
that it captures at least qualitatively, many of the features of the
effect of ripples on graphene transport.    

% Below we treat the
% static ripples by adapting a surface roughness model developed to treat the
% scattering of electrons off a spatially varying confining potential in
% Si-MOSFET devices~\cite{kn:ando1982}.  Whereas the formal validity of
% applying this surface roughness model to treat graphene ripples has yet
% to be established, we are persuaded that it provides the zeroth-order
% approximation.
          
\section{Formalism}

We calculate the Drude-Boltzmann conductivity (see 
Ref.~\cite{kn:adam2007a} for a detailed discussion on the range of 
validity of this formalism) using $\sigma = (2e^2/h)~E_{\rm F} \tau$,
\begin{eqnarray} 
% \sigma &=& \frac{2 e^2}{h} E_{\rm F} \tau, \nonumber \\
\tau^{-1} &=& \frac{4 E_{\rm F}}{\pi \gamma^2} 
\int_0^1 d\eta \ \eta^2 \sqrt{1-\eta^2} 
\left|\frac{v(\eta)}{\epsilon(\eta)}\right|^2,
\end{eqnarray}
where $E_{\rm F}$ is the Fermi energy, $\hbar^{-1}\gamma$ is the Fermi 
velocity and the momentum transfer $q = |{\bf k} - {\bf k}'| = 
2 k_{F} \eta$.  The different types of impurities change the
scattering potential $v(\eta)$ and screening is accounted
for using the RPA dielectric function $\epsilon(\eta)$ reported in 
Ref.~\cite{kn:hwang2006b}.  For Coulomb scatterers, we use 
\begin{equation}
v^{C}(\eta) = \frac{\sqrt{n_{\rm imp}}\pi e^2}{\kappa k_{\rm F} \eta},
\ \sigma^{C} = \frac{2 e^2}{h} \left( \frac{n}{n_{\rm imp}} \right)
    \frac{1}{F_1(2 r_s)},
\end{equation}
% and find~\cite{kn:adam2007a} 
% \begin{equation}
% \sigma^{C} = \frac{2 e^2}{h} \left( \frac{n}{n_{\rm imp}} \right)
%     \frac{1}{F_1(2 r_s)},
% \end{equation}
where $F_1(x)$ is given analytically by~\cite{kn:adam2007a}
\begin{equation}
\frac{F_1(x)}{x^2} = \frac{\pi}{4} + 3x -\frac{3x^2 \pi}{2}
   + x (3x^2 -2) \frac{\arccos[1/x]}{\sqrt{x^2 -1}}.
\end{equation}

It is straight forward to generalize this formalism to calculate the 
conductivity of screened short-range scatterers.  We have 
\begin{eqnarray}
v^{S}(\eta) &=& \sqrt{n_{\rm imp}} u, \ \ 
\sigma^{S} = \frac{4 \pi e^2}{h} \frac{\gamma^2}{n_{\rm imp} u^2 }
\frac{1}{F_2(2 r_s)}, \nonumber \\
F_2(x) &=& \frac{\pi}{2} - \frac{16 x}{3} + 40x^3 + 6 \pi x^2 
- \nonumber \\ && \mbox{}
20 \pi x^4 + 8 x^2 (5x^3 - 4x) \frac{\arccos[1/x]}{\sqrt{x^2 -1}}.
\end{eqnarray}

This result demonstrates that screening does not change the 
carrier density independence of the conductivity arising from 
short-range scattering and further predicts how the conductivity 
depends on the dielectric constant of the substrate.

In similar fashion, we can solve the surface roughness 
model~\cite{kn:ando1982}.  First to determine its importance near the 
Dirac point, we calculate the limit $n \rightarrow 0$ and 
find
\begin{eqnarray}
\sigma^R(n\rightarrow 0) = \frac{e^2}{4 \pi^2 h} 
\frac{1}{n^2 \Delta^2 \xi^2 r_s^2 F_2(2 r_s)} \sim \frac{1}{n^2}.
\end{eqnarray}

Therefore, within this surface roughness model ( and in a manner
qualitatively similar to short-range scattering), this effect becomes
unimportant at the lowest carrier densities and at low temperature.  
Since experimentally, the
parameter $\sqrt{\pi n^*} \xi \gtrsim 3$, the limit $k_{\rm F} \xi \gg
1$ is more realistic.  An analytic result can be obtained in the
unscreened limit (where we note that screening would only reduce the
contribution from this effect).  We find \begin{eqnarray} \ell^{R} &=&
\frac{\xi}{16 \sqrt{\pi} k_{\rm F}^2 r_s^2 \Delta^2}, \nonumber \\
\sigma^{R}(k_{\rm F} \xi \gg 1) &=& \frac{e^2}{h} \frac{\xi}{8
\sqrt{\pi} k_{\rm F} \Delta^2 r_s^2}, \end{eqnarray} where the first
equation suggests that the mean free path $\ell$ for surface
roughness scattering is typically larger
than the sample size.  These results, naturally, strongly depend on how
one models the surface roughness.  Recently, an alternate model proposed
by Ref.~\cite{kn:katsnelson2007} suggested that in some special
circumstances ripples can mimic Coulomb scatterers, although their
phonon model additionally predicts strong temperature dependence (above
a certain quenching temperature of about $100 {\rm K}$) -- an effect
that has not been observed in the experiments.  We do not believe that
the ripples in graphene are playing an important direct quantitative
role in determining the carrier mobility in currently available samples,
but rather scattering due to unintentional charged impurities is the
dominant scattering mechanism in graphene. 

\begin{figure}[h]
%h=here, t=top, b=bottom, p=separate figure page
\begin{center}\leavevmode
\includegraphics[width=0.9\linewidth]{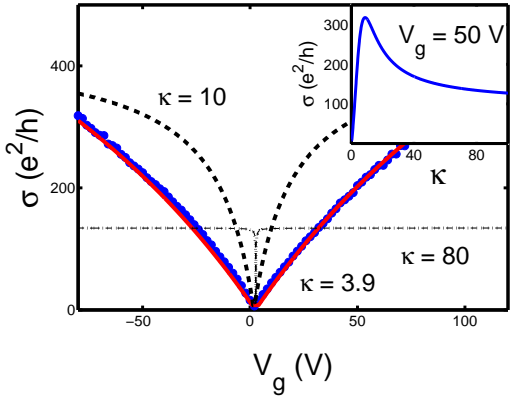}
\caption{Figure showing the dependence of conductivity on
substrate dielectric constant $\kappa$. Filled blue circles show
experimental data of Ref.~\protect{\cite{kn:cho2007}}.  Solid,
dashed and broken lines show theoretical results for $\kappa = 3.9,
10, 80$ respectively.  For large $\kappa$ short-range scattering
dominates while for small $\kappa$ Coulomb scattering dominates.  The
inset shows that for a fixed gate voltage, the conductivity has a
non-monotonic dependence on $\kappa$ which is a consequence of the
competition between short-range and long-range scattering.} 
\label{MainFigure}
\end{center}
\end{figure}

In Fig.~\ref{MainFigure} we show the effect of changing substrate
dielectric constant on graphene transport properties.  The blue
circles show the experimental data of Ref.~\cite{kn:cho2007} for a
high mobility sample, and the red solid curve ($\kappa = 3.9$) uses
the experimentally reported values in that paper for the undetermined
parameters in our theory.  A single fit parameter is still necessary
to obtain the overall scale of the short-range impurities.  Shown also
are theoretical predictions for different dielectric constants.  For
$\kappa \lesssim 5$, Coulomb scattering dominates (as is the case in
most current experimental samples), while for larger dielectric
constant, Coulomb scattering is more strongly screened making
short-range scattering more dominant (see broken curve with $\kappa =
80$).  For intermediate values of dielectric constant, we have a
crossover from low-density Coulomb-scattering-dominated behavior to
high-density short-range-scattering dominated (see dashed curve with
$\kappa = 10$).    
                  
\section{Discussion and Conclusion}

We have examined three different sources of scattering in graphene and 
treated them in the RPA-Boltzmann regime which we believe provides
the basic formalism to understand current graphene experiments.  This 
formalism strongly indicates that close to the Dirac point, the
transport properties are dominated by Coulomb impurity scattering.

The basic picture is quite intuitive.  As described above, within the
RPA-Boltzmann formalism, charged impurities in the substrate provide a
contribution to the conductivity that is proportional to the number of
carriers in the graphene sheet.  Additionally, these same charged
impurities generate potential fluctuations that are screened by any
carriers in the graphene sheet.  A self-consistent
procedure~\cite{kn:adam2007a} is used to determine the residual carriers
that contribute to Boltzmann conductivity even as the external gate
voltage is tuned through the duality point where the majority of
carriers changes from electrons to holes.  This mean-field residual
carrier density $n^*$ provides for a minimum conductivity plateau and is
in quantitative agreement with recent experimental
data~\cite{kn:tan2007, kn:cho2007}.  In this context, the current work
demonstrates that other scattering mechanisms do not alter this picture. 
Additionally, this work makes several predictions that can be tested
experimentally if one is able to change the substrate onto which graphene
has been cleaved.  Experiments have already been performed on graphene
sandwiched between various kinds of plastics using a transfer printing
method~\cite{kn:chen2007}  and experiments in the near future will 
certainly test the predictions we make in this work.

\par

This work is supported by US-ONR.

% \vspace{-0.1in}
% \bibliography{/home/sadam1/LaTex/shaffiquebib.bib}

\begin{thebibliography}{10}
\expandafter\ifx\csname url\endcsname\relax
  \def\url#1{\texttt{#1}}\fi
\expandafter\ifx\csname urlprefix\endcsname\relax\def\urlprefix{URL }\fi

\bibitem{kn:tan2007}
Y.-W. Tan, Y.~Zhang, K.~Bolotin, Y.~Zhao, S.~Adam, E.~H. Hwang, S.~\mbox{Das
  Sarma}, H.~L. Stormer, P.~Kim, Measurement of scattering rate and minimum
  conductivity in graphene, Preprint(arXiv:0707.1807v1 [cond-mat.mes-hall]).

\bibitem{kn:cho2007}
S.~Cho, M.~S. Fuhrer, Charge transport and inhomogeneity near the charge
  neutrality point in graphene, Preprint (arXiv:0705.3239v2
  [cond-mat.mes-hall]).

\bibitem{kn:hwang2006c}
E.~H. Hwang, S.~Adam, S.~\mbox{Das Sarma}, Carrier transport in 2d graphene
  layers, Phys. Rev. Lett. 98 (2007) 186806--186809.

\bibitem{kn:martin2007}
J.~Martin, N.~Akerman, G.~Ulbricht, T.~Lohmann, J.~H. Smet, K.~\mbox{von
  Klitzing}, A.~Yacobi, Observation of electron-hole puddles in graphene using
  a scanning single electron transistor, Preprint (arXiv:0705.2180v1
  [cond-mat.mes-hall]).

\bibitem{kn:dassarma2005b}
S.~\mbox{Das Sarma}, M.~P. Lilly, E.~H. Hwang, L.~N. Pfeiffer, K.~W. West,
  J.~L. Reno, Two-dimensional metal-insulator transition as a percolation
  transition in a high-mobility electron system, Phys. Rev. Lett. 94~(13)
  (2005) 136401--136405.

\bibitem{kn:adam2007a}
S.~Adam, E.~H. Hwang, V.~M. Galitski, S.~{\mbox Das Sarma}, A self-consistent
  theory for graphene transport, Proc. Natl. Acad. Sci. USA, in press
  (arXiv:0705.1540 [cond-mat.mes-hall]).

\bibitem{kn:cheianov2007}
V.~Cheianov, V.~Falko, B.~Altshuler, I.~Aleiner, Random resistor network model
  of minimal conductivity in graphene, Preprint (arXiv:0706.2968v2
  [cond-mat.mes-hall]).

\bibitem{kn:ishigami2007}
M.~Ishigami, J.~H. Chen, W.~G. Cullen, M.~S. Fuhrer, E.~D. Williams, Atomic
  structure of graphene on \mbox{$Si0_2$}, Nano Lett. 7 (2007) 1643.

\bibitem{kn:meyer2007}
J.~C. Meyer, A.~K. Geim, M.~I. Katsnelson, K.~S. Novoselov, T.~J. Booth,
  S.~Roth, The structure of suspended graphene sheets, Nature 446 (2007)
  60--63.

\bibitem{kn:ando2006}
T.~Ando, Screening effect and impurity scattering in monolayer graphene, J.
  Phys. Soc. Jpn. 75 (2006) 074716--074723.

\bibitem{kn:aleiner2006}
I.~Aleiner, K.~Efetov, Effect of disorder on transport in graphene, Phys. Rev.
  Lett. 97 (2006) 236801--236805.

\bibitem{kn:nomura2007}
K.~Nomura, A.~H. MacDonald, Quantum transport of massless dirac fermions in
  graphene, Phys. Rev. Lett. 98 (2007) 076602--076606.

\bibitem{kn:hwang2006b}
E.~H. Hwang, S.~\mbox{Das Sarma}, Dielectric function, screening and plasmons
  in 2d graphene, Phys. Rev. B 75~(20) (2007) 205418--206424.

\bibitem{kn:cheianov2006}
V.~V. Cheianov, V.~I. Fal'ko, Friedel oscillations, impurity scattering and
  temperature dependence of resistivity in graphene, Phys. Rev. Lett. 97 (2006)
  226801--226805.

\bibitem{kn:morozov2006}
S.~V. Morozov, K.~S. Novoselov, M.~I. Katsnelson, F.~Schedin, L.~A.
  Ponomarenko, D.~Jiang, A.~K. Geim, Strong suppression of weak localization in
  graphene, Phys. Rev. Lett. 97~(1) (2006) 016801.

\bibitem{kn:neto2007}
A.~H. \mbox{Castro Neto}, E.-A. Kim, Charge inhomogeneity and the structure of
  graphene sheets, Preprint (arXiv:0702562 [cond-mat.other]).

\bibitem{kn:katsnelson2007}
M.~I. Katsnelson, A.~K. Geim, Electron scattering on microscopic corrugations
  in graphene, Preprint (arXiv:0706.2490 [cond-mat.mes-hall]).

\bibitem{kn:ando1982}
T.~Ando, A.~B. Fowler, F.~Stern, Electronic properties of two-dimensional
  systems, Rev. Mod. Phys. 54 (1982) 437--672.

\bibitem{kn:chen2007}
J.~H. Chen, M.~Ishigami, C.~Jang, M.~Fuhrer, D.~Hines, E.~D. Williams, Printed
  graphene circuits, Adv. Mater. (2007) in press.

\end{thebibliography}
% \bibliographystyle{elsart-num}

\end{document}